# SDNGuardStack: An Explainable Ensemble Learning Framework for High-Accuracy Intrusion Detection in Software-Defined Networks

Ashikuzzaman
Dept. of CSE
University of Barishal
Barishal, Bangladesh
ashikuzzaman.cse7.bu@gmail.com

Mahabubur Rahman
Dept. of CSE
University of Scholars
Dhaka, Bangladesh
rahmanmahabubur771@gmail.com

Md. Mehedi Hasan
Dept. of CSE
University of Barishal
Barishal, Bangladesh
mmehedi20.cse@bu.ac.bd

Md. Saifuzzaman Abhi
Dept. of CSE
University of Barishal
Barishal, Bangladesh
msaifuzzaman20.cse@bu.ac.bd

Dr. Md. Manjur Ahmed*
Dept. of CSE
University of Barishal
Barishal, Bangladesh
mmahmed@bu.ac.bd

Md. Ahsan Arif
Dept. of CSE
University of Scholars
Dhaka, Bangladesh
ahsan@ius.edu.bd

*Corresponding Author

*Abstract*—Software-Defined Networking (SDN) is another technology that has been developing in the last few years as a relevant technique to improve network programmability and administration. Nonetheless, its centralized design presents a major security issue, which requires effective intrusion detection systems. The SDN-specific machine learning-based intrusion detection system described in this paper is innovative because it is trained and tested on the InSDN dataset which models attack scenarios and realistic traffic patterns in SDN. Our approach incorporates a comprehensive preprocessing pipeline, feature selection via Mutual Information, and a novel ensemble learning model, SDNGuardStack, which combines multiple base learners to enhance detection accuracy and efficiency. In addition, we include explainable AI methods, including SHAP to add transparency to model predictions, which helps security analysts respond to incidents. The experiments prove that SDNGuardStack has an accuracy rate of 99.98% and a Cohen Kappa of 0.9998, surpassing other models, and at the same time being interpretable and practically executable. It is interesting to see such features like Flow ID, Bwd Header Len, and Src Port as the most important factors in the model predictions. The work is a step towards closing the gap between performance intrusion detection and realistic deployment in SDN, which will lead to the creation of secure and resilient network infrastructures.

## I. INTRODUCTION

Software-Defined Networking (SDN) is a relatively young technology that, within recent years, became one of the foundations of programmable networks and traffic control with centralized management [1], [2]. Since SDN will expand its use in enterprise and cloud networks, which is projected to increase to USD 70 billion by 2030, it also creates additional avenues of attack because of its logically centralized design. As the control hub, the SDN controller introduces a serious point of vulnerability [3]. Such compromise has the potential to interfere with or divert behavior on a network-wide basis. Conventional intrusion detection systems (IDS), many of which are signature-based, do not scale to SDN environments because they are static and cannot be used to identify threats that are dynamic [4]. Machine learning (ML) based approaches represent an attractive alternative as they can identify traffic patterns and extrapolate to new attacks [5]–[9]. Nevertheless, most current SDN-oriented ML approaches suffer significant drawbacks: high misclassification rate, lack of explainability and high computational costs that make real-time deployment impractical [10] . Recent benchmark datasets (NSL-KDD, CICIDS2017, UNSW-NB15) do not reflect well the current SDN traffic behavior. The InSDN dataset overcomes such a shortcoming by offering realistic, SDN-specific traffic flows and current attack classes. It allows the researchers to investigate ML-based detection under a more realistic setting. The proposed work is an ML-based intrusion detection system specifically designed to work with SDN, with the InSDN dataset used to give the work a more real-world application. The methodology does not only consider the accuracy of detection, but it is also efficient and explainable, which guarantees its practical application to real-time SDN systems.

The key contributions of this work are:

- An in-depth preprocessing pipeline including noise filtering, feature selection, and class balancing is implemented to improve data quality and model generalizability.
- Multiple ML models are benchmarked and optimized to enhance intrusion detection accuracy across diverse attack categories.
- The proposed architecture minimizes computational complexity to ensure compatibility with SDN controller runtime environments.
- A novel ensemble learning model is introduced, combining base learners to achieve higher robustness and predictive stability.
- Explainable AI techniques such as SHAP and LIME are integrated, offering transparency into model decisions and aiding security analysts in incident response.

All these contributions are meant to close the gap between high-performance intrusion detection and realistic SDN deployment.

The remainder of the document is organized as follows: Section 2 contains the relevant work on SDN-based intrusion detection with ML is reviewed. Section 3 provides information on the dataset, preprocessing steps and overall methods. Section 4 discusses evaluation metrics and experimental results. Section 5 summarizes the paper and suggests topics of future research.

## II. RELATED WORK

Machine learning solutions to intrusion detection in SDNs have been varied with high accuracy as the goal. However, it is challenging to find a compromise between detection performance, interpretability and computational efficiency. In this section, the literature that considers these aspects will be reviewed.

Ataa et al. [11] proposed a CNN-LSTM and Transformer-based hybrid intrusion detection framework for the dataset InSDN. Strong detection performance was attained by their model, with the Transformer component attaining 99.02% accuracy in binary classification tasks. Although the method demonstrated high generalization across attack types, it lacked interpretability features such as SHAP or LIME, which are essential for understanding decision rationale in security contexts. Additionally, Transformer models incur heavy computational overhead, making them unsuitable for real-time SDN deployment in latency-sensitive environments. Ibrahim et al. [12] developed a classical machine learning-based approach using Random Forest and KNN classifiers along with SMOTE to handle class imbalance. Their pipeline surpassed 99% accuracy on the InSDN dataset and performed well across diverse attack classes. However, the absence of interpretability mechanisms and reliance on manual preprocessing and feature engineering introduce operational complexity. These factors, coupled with limited model transparency, hinder its effectiveness in dynamic and real-time SDN environments. Shariff et al. [13] proposed a deep learning framework leveraging LSTM networks optimized with Bayesian tuning. Their system achieved up to 99.6% accuracy in detecting DDoS attacks, benefiting from LSTM's ability to capture sequential patterns in traffic. Despite its accuracy, the model lacks explainable AI (XAI) tools and requires computationally intensive Bayesian optimization, limiting its suitability for time-critical SDN controller operations. Mills et al. [14] introduced a hybrid intrusion detection and prevention framework utilizing ensemble stacking of multiple supervised classifiers. This approach outperformed individual models in accuracy, precision, and recall. However, it significantly increased model complexity, memory usage, and inference latency. The framework also omitted interpretability techniques, making it difficult for network administrators to trace prediction outcomes—thereby reducing operational trust and regulatory compliance. Wang et al. [15] applied Temporal Convolutional Networks (TCNs) for anomaly detection in SDNs, achieving a precision of 98.15%. TCNs provided a lighter alternative to RNNs for modeling sequential data. However, the approach lacked algorithmic and visual interpretability features and required considerable computational resources during training and inference. These limitations diminish its feasibility in resource-constrained or latency-sensitive SDN environments. Xu et al. [16] proposed the ESWO-IDM model, which integrates the Spider Wasp Optimizer with K-Nearest Neighbors (KNN) to enhance detection accuracy. The model achieved 98.83% accuracy, illustrating the benefit of metaheuristic-based feature selection. Nevertheless, both the optimizer and KNN lack interpretability, and KNN's high inference time complexity reduces scalability. These drawbacks make the approach unsuitable for real-time SDN intrusion detection systems. Fotse et al. [17] introduced FedLAD, a federated learning-based DDoS detection framework for large-scale SDNs. It attained approximately 98% accuracy while ensuring local data privacy across distributed nodes. Despite its privacy advantages, FedLAD suffers from communication and synchronization overhead that can increase latency. Furthermore, the absence of global interpretability tools limits transparency and hampers its adoption in real-time, operational SDN environments.

## III. METHODOLOGY

In this section, we describe the methodology of this research in detail. Figure 1 indicates the full pipeline of the proposed workflow, starting with data acquisition, explanatory intrusion detection based on ensemble learning and model interpretation methods.

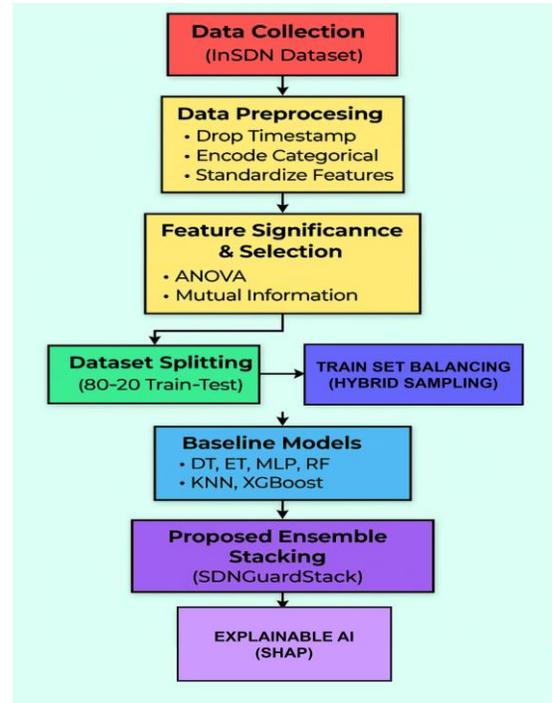

Fig. 1: The proposed workflow for SDN Intrusion detection.

### A. Data Collection

The work is conducted on the basis of the InSDN dataset, which is a realistic and SDN-specific intrusion detection dataset hosted on Kaggle [18], [19]. It consists of 343 889 traffic records, 83 features (79 numerical and 4 categorical (object)). The numerical features consist of different statistics of the flow level, including the number of packets, the byte rate, or the flow duration and the categorical features are the protocol types, IP addresses, or labels.

The dataset consists of 8 different kinds of attacks as well as benign (normal) traffic, which covers a large variety of threats relevant to SDN environments.

TABLE I: Attack Types in the InSDN Dataset

| Attack | Description | Count |
|---|---|---|
| Probe | Port/service scan | 98,129 |
| DDoS (1) | Distributed DoS variant 1 | 73,529 |
| Normal | Benign SDN traffic | 68,424 |
| DoS | Denial of Service | 53,616 |
| DDoS (2) | Distributed DoS variant 2 | 48,413 |
| BFA | Brute-force attempts | 1,405 |
| Web-Attack | Malicious HTTP requests | 192 |
| Botnet | C&C infected traffic | 164 |
| U2R | Privilege escalation | 17 |

## B. Preprocessing

The InSDN dataset was preprocessed using various techniques in order to make it ready to be seen by machine learning algorithms. The `Timestamp` feature was dropped because it did not have predictive value. Categorical variables such as `Dst IP`, `Src IP`, `Flow ID`, and `Label` were encoded with label encoding to change textual representation to numerical representation. This was to allow compatibility with standard ML algorithms. As there was high variance in numerical features, they were standardized by z-score normalization such that each feature has zero mean and unit variance. These measures mitigated the effect of scale and advanced the rate of convergence as well as the general learning performance of the models.

## C. Significance of Features using One-Way ANOVA

To ascertain feature relevance a One-Way ANOVA test was conducted. The variables `Tot Fwd Pkts` and `Subflow Fwd Pkts` were identified as insignificant having p-value more than 0.05. These characteristics showed insignificant differences across the various types of attacks and were not considered in the subsequent analysis to enhance model effectiveness.

## D. Feature Selection

To check the dependence between each feature and the target variable, Mutual Information (MI) was utilized. The `mutual_info_classif` scoring function along with `SelectKBest` was used to pick the top 15 features based on the MI scores. The features that are chosen have the greatest correlation with the categories of attacks and, therefore, enhance the interpretability of the model and decrease its dimensionality.

The relative importance of the top 15 features is demonstrated in Table II, which indicates the MI scores of all features.

TABLE II: Top 15 Features Selected by Mutual Information

| Feature | MI Score | Feature | MI Score |
| --- | --- | --- | --- |
| Flow ID | 1.5689 | Bwd IAT Max | 1.0322 |
| Bwd Header Len | 1.2519 | Bwd Pkts/s | 1.0084 |
| Src IP | 1.2220 | Flow IAT Mean | 1.0030 |
| Dst Port | 1.1343 | Flow Pkts/s | 1.0009 |
| Dst IP | 1.0474 | Init Bwd Win Byts | 0.9804 |
| Bwd IAT Mean | 1.0464 | Flow Duration | 0.9464 |
| Bwd IAT Tot | 1.0376 | Src Port | 0.9438 |
|  |  | Flow IAT Max | 0.9424 |

## E. Dataset Splitting

To balance model training and evaluation, the dataset was split 80/20 between training and testing sets. The split distribution of the dataset is provided in Table III.

TABLE III: Training and Testing Sample Counts per Class after 80-20 Split

| Class (Attack Type) | Training Samples | Testing Samples |
| --- | --- | --- |
| 2 (Probe) | 97,524 | 19,626 |
| 5 (DDoS) | 78,537 | 14,706 |
| 4 (Normal) | 54,823 | 13,685 |
| 3 (DoS) | 42,809 | 10,723 |
| 0 (BFA) | 1,123 | 281 |
| 7 (Web-Attack) | 149 | 38 |
| 1 (BOTNET) | 132 | 33 |
| 6 (U2R) | 14 | 3 |

*1) Dataset Balancing:* Since there was clear class imbalance in the training set, hybrid sampling method was used. Oversampling and undersampling were done on the majority classes to 30,000 samples each and minority classes to 30,000 samples. This class weighted scheme provides equal representation to all classes (30,000 samples per class) reducing the bias and resulting in a significant increase in the model performance on the underrepresented attack types.

## F. Baseline Models

A total of six popular machine learning models were trained to provide a baseline of performance of SDN intrusion detection. Decision Tree (DT) is a model with low complexity and interpretation which partitions data according to feature threshold. Extra Trees (ET) is a variant of Random Forest with random splits which tends to decrease variance. A feed-forward neural network called the Multi-Layer Perceptron (MLP) is capable of modeling intricate nonlinear interactions. Random Forest (RF) constructs a number of decision trees and averages their prediction to enhance precision and minimize overfitting. K-Nearest Neighbors (KNN) uses the most common label of the nearest neighbors to classify the samples, thus it is intuitive, however it is computationally expensive on large scale data. Lastly, the eXtreme Gradient Boosting (XGBoost) is an efficient and scalable boosting algorithm that has high predictive power and can regularize. These models have been chosen due to their variability in style and excellent performance in classification issues.

*1) SDNGuardStack: Proposed Ensemble Stacking Model:* Ensemble stacking model proposed in this research is aimed at combining several base learners to exploit the strengths of each learner and enhance the overall intrusion detection performance [20]. The base learners are Decision Tree (DT), Extra Trees (ET), and Multi-Layer Perceptron (MLP). The meta-learner receives the features of the predictions made by these models, which are independently learnt on the training data. The Light Gradient Boosting Machine (LGBM) classifier, which is effective in combining the predictions of the base models to produce the final prediction, is used to implement the meta-learner. The idea of this hierarchical approach is to achieve better detection performance bycovering a wide variety of decision boundaries and minimizing model bias. Pseudocode Algorithm 1 sums up the training procedure of the suggested stacking ensemble.

**Algorithm 1** Stacking Ensemble for SDN Intrusion Detection

**Require:** Training dataset $D = \{(x_i, y_i)\}_{i=1}^{N}$
**Require:** Base learners: Decision Tree (DT), Extra Trees (ET), MLP
**Require:** Meta learner: LightGBM (LGBM)
0: Split dataset $D$ into training set $D_{train}$ and validation set $D_{val}$
0: **for** each base learner $m$ in *{DT, ET, MLP}* **do**
0:   Train $m$ on $D_{train}$
0:   Generate predictions $P_m$ on $D_{val}$
0: **end for**
0: Construct new dataset $D'_{val}$ using predictions $\{P_{DT}, P_{ET}, P_{MLP}\}$ as features
0: Train meta learner LGBM on $D'_{val}$ with true labels from $D_{val}$
0: **Output:** Trained stacking ensemble model =0

## G. Explainable AI for Model Transparency

SHapley Additive exPlanations (SHAP) was applied to explain the model decisions measuring feature contributions [21], [22]. This approach gives appropriate answers on what characteristics are relevant to predictions to enhance the transparency and confidence in the intrusion detection system, which is crucial in SDN security applications.

## IV. DISCUSSION AND RESULTS

This part provides the experimental outcomes and performance comparison of the proposed SDNGuardStack model with some baseline classifiers. The models were evaluated with common classification metrics, accuracy, precision, recall, F1-score, and Cohens Kappa along with computational cost measures, like training time.

Table IV shows the comparison of the results of various machine learning models on training and testing data SDN intrusion detection. The capacities of all models to learn distinguishing features of the dataset are good, as they reach almost perfect accuracy, precision, recall, and F1-score on the training set. The proposed SDNGuardStack model significantly outperforms the baseline classifiers on the testing set and obtains the highest scores on all metrics (accuracy, precision, recall, and F1-score at 0.9998). This shows that it can generalize and detect intrusions in unseen data better, a crucial property of SDN security solutions that have to be used in real time. Other models like Decision Tree, Extra Trees, and Random Forest also achieve high results, yet their evaluating measures are slightly lower than SDNGuardStack, which implies that the stacking ensemble structure is not pointless in terms of gaining performance benefit. All models have minimal variations between training and testing results indicating that overfitting is sufficiently avoided. In order to give a more detailed analysis, Table V below shows the 5-fold cross-validation accuracy of the training data and Cohen Kappa scores of the testing data. These other measures provide additional information of the robustness, consistency and agreement of the models over and above accuracy and supports the soundness of the proposed SDNGuardStack in different network settings.

The results in Table V indicate the 5-fold cross-validation run on the validation accuracy and Cohen Kappa for the different models. The performance of all models is very high, as validation accuracy and kappa statistics are nearly equal to 1.0, which means that the models have high predictive ability. The suggested SDNGuardStack model provides the most accurate outcomes with the highest validation accuracy of 1.0000 and the maximum Cohen kappa value of 0.9998, indicating its overall capability to identify SDN intrusions most effectively. Other models such as Extra Trees, XGBoost, and Random Forest have also performed well but a little bit lower than SDNGuardStack. Such findings indicate the strength and efficiency of the SDNGuardStack model in precision and stability of intrusion detection in SDN networks; however, at the expense of greater computational expense during training as outlined in Table VI, which reports training times of the reviewed models.

The training time of every assessed model is stated in Table VI. The classical machine learning models also offer the shortest training time: K-Nearest Neighbors (1.12 seconds), but tree-based models, such as Extra Trees and Random Forest require more time (approx. 30 seconds). The neural network structure of the Multi-Layer Perceptron makes it time consuming (54.58 seconds). Proposed model SDNGuardStack offers the longest training time (70.45 seconds) since it is more extensive and intricate in its structure to ensure better detection capabilities. Even though SDNGuardStack requires increased computation resources in training, its high accuracy and strong performances pay off in the game of secure and reliable SDN intrusion detection.

The evaluation results of the proposed SDNGuardStack model are in Figure 2. The confusion matrix (top-left) indicates a great classification accuracy, especially on high frequency classes like Class 2, 3, and 5 with few misclassifications overall. This is additionally confirmed by the Precision-Recall curve (top-right) where the AUC values are close to perfect on most classes (AUC 0.9997), but the AUC values on Class 6 is much lower at 0.7238, suggesting that this class is comparatively difficult to isolate - perhaps because of class imbalance or feature overlap. The learning curve (bottom-left) illustrates a stable performance when the size of training set grows, the validation accuracy is gradually growing and almost coincides with the training accuracy, indicating a good generalization and a low degree of overfitting. The ROC curve (bottom-right) further supports the high detection power of the model, where all classes have almost perfect AUCs, meaning high true positives rates and low false positive rates. All together, these numbers confirm the soundness, basically pointing to the generalizability, and high precision of the suggested SDNGuardStack in SDN intrusion detection contexts.

### A. Feature Impact Detection via SHAP in SDNGuardStack

SHAP (SHapley Additive exPlanations) was used to assess feature importance in the multiple traffic classes to interpret the decisions made by the model. Overall contribution was found to be the highest in `Flow ID`, then `Bwd Header Len`, `Src Port` and `Src IP` (shown in Figure 3), which implies that these features have a great impact in differentiating benign and malicious SDN traffic. There was also some class-wise contributions such that `Bwd Header Len` contributed greatly to Classes 2 and 3 whereas `Src IP` contributed significantly to Class 5. Features such as `Bwd IAT Max` and `Bwd IAT Mean` have low SHAP values indicating they are not very relevant. The given analysis makes the suggested ensemble more interpretable because the globally and locally important features have been determined.

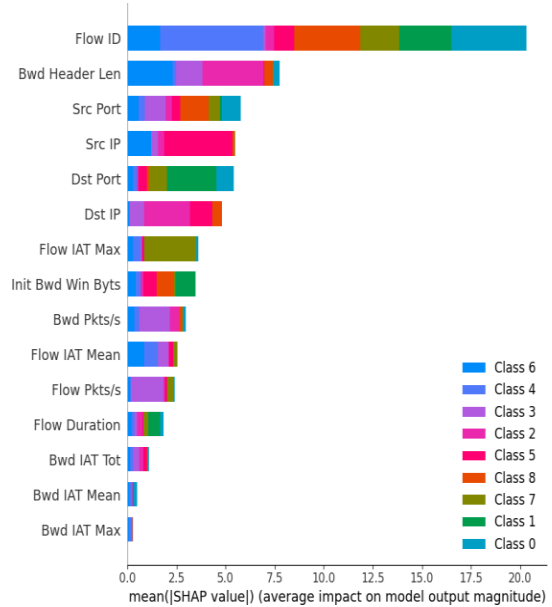

Fig. 3: SHAP summary plot showing the mean absolute SHAP value for top contributing features across SDN traffic classes.

### B. Comparative Analysis with previous studies

As can be seen in the comparative analysis in Table VII, some of the recent intrusion detection frameworks propose high accuracy results greater than 99%, but none of them considers explainable AI (XAI) methods to improve the transparency of the models. Not only does the proposed SDNGuardStack framework achieve the best accuracy of 99.98%, but it also takes into consideration XAI techniques, which introduce essential interpretability usually lacking in the previous works. Such a combination of excellent performance and explainability has made SDNGuardStack exceptionally well applicable in real-time secure and trustworthy Software-Defined Networking (SDN) operation, covering both detection performance and operational visibility.

TABLE IV: Performance Comparison of Models on Training and Testing Sets

| Model | Training | | | | Testing | | | |
|---|---|---|---|---|---|---|---|---|
| | Accuracy | Precision | Recall | F1-score | Accuracy | Precision | Recall | F1-score |
| Decision Tree [23] | 1.0000 | 1.0000 | 1.0000 | 1.0000 | 0.9997 | 0.9997 | 0.9997 | 0.9997 |
| Extra Trees [24] | 1.0000 | 1.0000 | 1.0000 | 1.0000 | 0.9997 | 0.9997 | 0.9997 | 0.9997 |
| K-Nearest Neighbors [25] | 0.9998 | 0.9998 | 0.9998 | 0.9998 | 0.9994 | 0.9994 | 0.9994 | 0.9994 |
| Multi-Layer Perceptron [26] | 0.9997 | 0.9997 | 0.9997 | 0.9993 | 0.9993 | 0.9993 | 0.9993 | 0.9993 |
| Random Forest [27] | 1.0000 | 1.0000 | 1.0000 | 1.0000 | 0.9997 | 0.9997 | 0.9997 | 0.9997 |
| XGBoost [28] | 1.0000 | 1.0000 | 1.0000 | 1.0000 | 0.9996 | 0.9996 | 0.9996 | 0.9996 |
| Proposed SDNGuardStack | 1.0000 | 1.0000 | 1.0000 | 1.0000 | **0.9998** | **0.9998** | **0.9998** | **0.9998** |

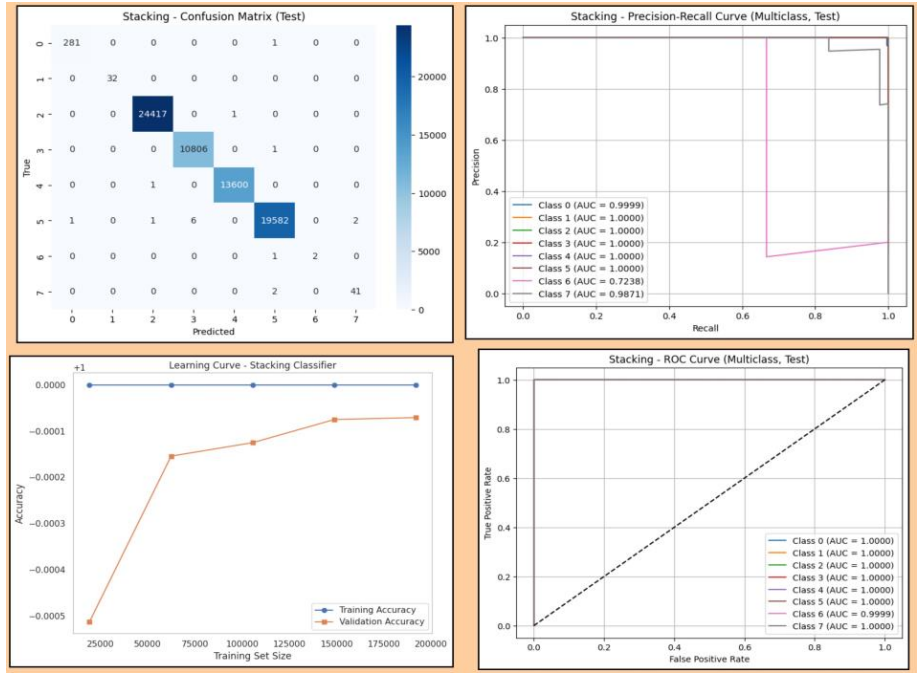

Fig. 2: Comprehensive evaluation results of the SDNGuardStack model: confusion matrix (top-left), precision-recall curve (top-right), learning curve (bottom-left), and ROC curve (bottom-right).

TABLE V: 5-Fold Cross-Validation: Validation Accuracy and Cohen's Kappa

| Model | Validation Accuracy | Cohen's Kappa |
|---|---|---|
| Decision Tree | 0.9998 | 0.9996 |
| Extra Trees | 0.9999 | 0.9997 |
| K-Nearest Neighbors | 0.9997 | 0.9992 |
| Multi-Layer Perceptron | 0.9996 | 0.9991 |
| Random Forest | 0.9999 | 0.9996 |
| XGBoost | 0.9999 | 0.9997 |
| Proposed SDNGuardStack | **1.0000** | **0.9998** |

TABLE VI: Training Time of Different Models (in seconds)

| Model | Training Time (s) |
|---|---|
| Decision Tree | 2.85 |
| Extra Trees | 29.47 |
| Random Forest | 32.96 |
| K-Nearest Neighbors | 1.12 |
| Multi-Layer Perceptron | 54.58 |
| XGBoost | 17.58 |
| Proposed SDNGuardStack | 70.45 |

V. CONCLUSION

In this work, we propose SDNGuardStack, an effective machine learning-based intrusion detector system tailored to Software-Defined Networking (SDN) scenarios. Using InSDN dataset, the model dis-plays a remarkable accuracy of 99.98% and a Cohen Kappa statistic of 0.9998, also identifying different attack scenarios effectively. The model can be improved with the explainable AI methods (SHAP and LIME), which increase the transparency and enable security analysts to see the contribution of features, with the most important features being Flow ID, Bwd Header Len and Src Port, reported as important predictors. The work has addressed the most important issues in SDN security, and also it offers a realistic solution that can be implemented in real time.

TABLE VII: Comparison of Intrusion Detection Studies: Model, Accuracy, and XAI Support

| Study | Model | Accuracy (%) | XAI Support |
|---|---|---|---|
| [11] | CNN-LSTM | 99.02 | No |
| [12] | Random Forest | >99 | No |
| [13] | LSTM | 99.60 | No |
| [14] | Stacking Classifiers | >99 | No |
| **This Study** | **SDNGuardStack** | **99.98** | **Yes** |

In future work, the scaling of SDNGuardStack will be optimized, which will allow working with bigger datasets and in more intricate network settings. cutting-edge explainability techniques will also be discussed, to expand on clarifying model decisions further, getting the users to trust it more. Furthermore, incorporation of real-time data streams will enable adaptive learning, whereby the model can keep on enhancing its detection ability based on the changing threats. The work is done to secure and strengthen the SDN infrastructures to meet the emerging cyber threats, and assure their integrity and reliability in the dynamic network environment. This work is part of the current research to effectively secure SDN systems.